\documentclass[12pt]{article}
\def\beq{\begin{equation}}
\def\eeq{\end{equation}}
\def\ap#1#2#3 {Ann. Phys. (NY) {\bf#1} (19#2) #3}
\def\err#1#2#3 {{\it Erratum} {\bf#1} (19#2) #3}
\def\ib#1#2#3 {{\it ibid.} {\bf#1} (19#2) #3}
\def\ijmp#1#2#3 {Int. J. Mod. Phys. {\bf#1} (19#2) #3}
\def\jetp#1#2#3 {JETP Lett. {\bf#1} (19#2) #3}
\def\mpl#1#2#3 {Mod. Phys. Lett. {\bf#1} (19#2) #3}
\def\np#1#2#3 {Nucl. Phys. {\bf#1} (19#2) #3}
\def\pl#1#2#3 {Phys. Lett. {\bf#1} (19#2) #3}
\def\prep#1#2#3 {Phys. Rep. {\bf#1} (19#2) #3}
\def\prev#1#2#3 {Phys. Rev. {\bf#1} (19#2) #3}
\def\prl#1#2#3 {Phys. Rev. Lett. {\bf#1} (19#2) #3}
\def\sjnp#1#2#3 {Sov. J. Nucl. Phys. {\bf#1} (19#2) #3}
\def\spj#1#2#3 {Sov. Phys. JETP {\bf#1} (19#2) #3}
\def\spu#1#2#3 {Sov. Phys. Usp. {\bf#1} (19#2) #3}
\def\zp#1#2#3 {Zeit. Phys. {\bf#1} (19#2) #3}
\def\ecp {\eta_c^\prime}

\begin{document}
\begin{titlepage}
\begin{center}
{\Large \bf Theoretical Physics Institute \\
University of Minnesota \\}  \end{center}
\vspace{0.2in}
\begin{flushright}
TPI-MINN-02/20-T \\
UMN-TH-2105-02 \\
June 2002 \\
\end{flushright}
\vspace{0.3in}
\begin{center}
{\Large \bf  The decay $\eta_c^\prime \to \eta_c \, \pi \, \pi$.
\\}
\vspace{0.2in}
{\bf M.B. Voloshin  \\ }
Theoretical Physics Institute, University of Minnesota, Minneapolis,
MN
55455 \\ and \\
Institute of Theoretical and Experimental Physics, Moscow, 117259
\\[0.2in]
\end{center}

\begin{abstract}
It is pointed out that the decay of the recently observed charmonium
$\eta_c^\prime$ resonance, $\eta_c^\prime \to \eta_c \, \pi \, \pi$ is
simply related to the well studied decay $\psi^\prime \to J/\psi \, \pi
\, \pi$ and can thus be used for absolute normalization of other decay
modes of the $\eta_c^\prime$. The total rate of the discussed decay
should be approximately three to four times the corresponding rate for
the $\psi^\prime$ resonance making the channel with charged pions the
most probable exclusive decay mode of the $\eta_c^\prime$ with the
branching ratio in the range 5-10\%.
\end{abstract}

\end{titlepage}

The recent observation\cite{belle} of the charmonium state $\ecp$ with
mass $3654\pm 6 (stat) \pm 8 (syst) \, MeV$ illustrates the possibility
of studying charmonium resonances produced in the decays of $B$
mesons\cite{elq}. Despite being known for more than a quarter of a
century the charmonium system still leaves an ample space for further
study, especially in the spin singlet sector (a recent update on the
singlet states is given in Ref.~\cite{gr}). Even for the most well
established singlet state $\eta_c$ the data on its properties are still
quite scarce, and its total decay width is measured with a quite
`modest' accuracy $13.2^{+3.8}_{-3.2} \, MeV$\cite{pdg}, which is hardly
improved by the recent CLEO result of $26 \pm 6 \, MeV$\cite{cleo}. For
the newly observed $\ecp$ it is unlikely that a direct measurement of
its total width can be any easier. The  purpose of the present letter is
to point out that in the case of $\ecp$ the differential spectrum and
the total rate of the decay $\eta_c^\prime \to \eta_c \, \pi \, \pi$,
first discussed in Ref.~\cite{ok}, can be found with absolute
normalization through its relation to the well studied decay
$\psi^\prime \to J/\psi \, \pi \, \pi$. As simple as this remark is, it
still merits being explicitly made, since this decay can provide a
useful normalization for other decay channels of $\ecp$ as well as for
the total decay rate, at least in the interim, until
$\Gamma_{tot}(\ecp)$ is measured directly. Moreover, due to the known
(and, to an extent, understood) behavior of the decay amplitude, a
slightly larger energy release in the transition between the singlet
states is quite important, so that the total rate $\Gamma(\eta_c^\prime
\to \eta_c \, \pi \, \pi)$ should be three to four times larger than
that for the transition between the vector states. In absolute numbers
this corresponds to $\Gamma(\eta_c^\prime \to \eta_c \, \pi^+ \, \pi^-)
\approx  2 \Gamma(\eta_c^\prime \to \eta_c \, \pi^0 \, \pi^0) \approx
300 \pm 50 \, KeV$\footnote{The actual range of uncertainty is somewhat
difficult to estimate, since it includes both the experimental errors in
$\Gamma(\psi^\prime \to J/\psi \, \pi \, \pi)$ and in the mass of
$\ecp$, as well as the theoretical uncertainty of the $v^2/c^2$ terms in
the non-relativistic expansion. Given the number of contributing
factors, one should rather treat the quoted error bars as an estimate of
``$1 \, \sigma$".} which should constitute 5-10\% of the expected (with
substantial uncertainty) total width of the $\ecp$. Clearly, this
implies that the discussed pionic transitions are the most probable
exclusive decay channels of the $\ecp$, much in the same way as they are
for the $\psi^\prime$.

The outline of the reasoning is as follows.
Hadronic transitions in heavy quarkonium are described in terms of the
multipole expansion in QCD\cite{kg,mv}. Within this expansion the
amplitudes of the decays $\eta_c^\prime \to \eta_c \, \pi \, \pi$ and
$\psi^\prime \to J/\psi \, \pi \, \pi$ are exactly equal in the
non-relativistic limit for heavy quarks. Furthermore the amplitude for
emission of the pions by the relevant gluonic operator is governed by
the current algebra and the trace anomaly in QCD\cite{vz,ns}. The
presence of the anomaly contribution substantially enhances the
amplitude of these decays and gives rise to a rapid growth of the
amplitude with the invariant mass of the dipion.

Proceeding to a more detailed discussion, we start with the remark that
the emission of light hadrons in transitions between heavy quarkonium
levels goes through quarkonium interaction with soft gluon fields which,
in turn, emit the light mesons. Since at large quark mass the heavy
quarkonium is (at least formally) a compact object, its interaction with
soft gluon field can be expanded in multipoles\cite{kg,mv}. The leading
term in the expansion is the $E1$ interaction with the chromo-electric
component ${\bf E}^a$ of the gluonic field strength tensor. The
effective Hamiltonian for this interaction has the standard form
\beq
H_{E1}=-{1 \over 2} \, g \, \xi^a \, ({\bf r} \cdot {\bf E}^a)~,
\label{he1}
\eeq
where $g$ is the QCD coupling, $\xi^a=t_1^a-t_2^a$ is the difference of
the color generators acting on the heavy quark and antiquark
respectively, and ${\bf r}={\bf r}_1-{\bf r}_2$ is the relative position
of the quark and the antiquark.

The emission of two pions in transitions between $S$ states proceeds in
the second order in the $E1$ interaction. Since the heavy quark spin is
not involved in this interaction, and since it also factorizes in the
quarkonium wave function in the non-relativistic limit, the amplitude of
the two pion emission does not depend on the heavy quark spin and thus
is the same for transitions between spin-triplet and spin-singlet
states. The expression for the amplitude can be written as\cite{vz,vyz}
\beq
A_{\pi\pi}=\langle \pi \pi | \pi \alpha_s \, ({\bf E}^a \cdot {\bf E}^a)
| 0 \rangle \, A_0~,
\label{app}
\eeq
where $A_0$ is the quarkonium transition matrix element
\beq
A_0={1 \over 24} \,\langle 1S | \xi^a \, r_i \, {\tilde G} \, r_i \,
\xi^a | 2S \rangle={2 \over 9} \,\langle 1S |  r_i \, {\tilde G} \, r_i
\, | 2S \rangle~,
\label{a0}
\eeq
with ${\tilde G}$ being the Green function for the quark-antiquark pair
in the color octet state.

The matrix element for the production of two pions by the gluonic
operator $\alpha_s \, ({\bf E}^a \cdot {\bf E}^a)$ in eq.(\ref{app}) has
been understood\cite{vz,ns} through its relation to the conformal
anomaly in QCD and the current algebra. Namely, one can write this
operator as
\beq
\alpha_s \, ({\bf E}^a \cdot {\bf E}^a) = -{\alpha_s \over 4} \, \left(
F_{\mu \nu}^a \right )^2 + {\alpha_s \over 2} \left ( ({\bf E}^a)^2  +
({\bf B}^a)^2 \right )={2 \pi \over b} \, \theta_\mu^\mu + \alpha_s \,
\theta_{00}^G~,
\label{theta}
\eeq
with $\theta_{\mu \nu}$ being the energy-momentum tensor in QCD and
$\theta_{\mu \nu}^G$ its gluonic part. The first term in the last
expression arises from the trace anomaly, i.e. in the chiral limit one
has $\theta_\mu^\mu=-(b \, \alpha_s/8 \pi) \, (F_{\mu \nu}^a)^2$, where
$b=3$ is the coefficient in the QCD beta-function with three quarks.

Finally, the matrix element of $\theta_\mu^\mu$ over the pions is
governed by a current algebra low-energy theorem\cite{vz}, which in the
chiral limit yields, e.g. for the pair of charged pions
\beq
\langle \pi^+ \pi^- | \, \theta_\mu^\mu \, | 0 \rangle =q^2~,
\label{let}
\eeq
where $q=p_++p_-$ is the total 4-momentum of the pion pair. Therefore
the two pion transition amplitude (\ref{app}) can be parameterized as
\beq
A_{\pi^+ \pi^-}={2 \pi^2 \over b}\, (q^2-C) \, A_0~.
\label{aparam}
\eeq
The term $C$ stands here for the terms that are smaller than the anomaly
contribution in the chiral limit, namely for the terms of order
$m_\pi^2$ and those coming from $\alpha_s \, \theta_{00}^G$ in
eq.(\ref{theta}). The latter contribution\cite{ns}, formally, is not
vanishing in the chiral limit and thus is not parametrically of order
$m_\pi^2$. Neither it is a constant, independent of $q^2$. However, it
is numerically small and is comparable to terms of order $m_\pi^2$ and
also it varies over the physical region of $q^2$ sufficiently slowly, so
that the effect of this variation can be neglected.

The parameterization (\ref{aparam}), first considered within the general
framework of chiral symmetry in Refs.~\cite{mv0,bc}, with a small and
constant $C$ exceptionally well describes the dipion invariant mass
spectrum in the transitions from $\psi^\prime$ and $\Upsilon^\prime$,
where no deviation from this parameterization has been observed so far
(see e.g. a discussion in the review \cite{vyz}).\footnote{This
parameterization fails to correctly describe the observed behavior in
the decay $\Upsilon(3S) \to \Upsilon \, \pi \, \pi$, which can be caused
by several reasons, generally not directly related to the transitions
between $2S$ and $1S$ states. For a discussion see Ref.~\cite{vyz} and
also \cite{ckk}.} Numerically the fit from the data gives $C=(4.6\pm
0.2) \, m_\pi^2$ for the $\psi^\prime$ decay, and $C=(3.3 \pm 0.2) \,
m_\pi^2$ for the $\Upsilon^\prime$ decay. (It can be noted that a
smaller value of $C$ in the latter decay than in the former had in fact
been predicted\cite{ns} well before the data became available.)

The absence of dependence of the amplitude in eq.(\ref{app}) on the
heavy quark spin is generally broken by the spin-spin and tensor
interaction between the quark and antiquark, which result in terms of
order $v^2/c^2$ in the non-relativistic expansion. For the $2S$ and $1S$
states of charmonium such terms are generally estimated at the level of
15-20\%. Thus, up to such uncertainty one can consider the amplitude of
the decay $\ecp \to \eta_c \, \pi \, \pi$ as being equal to that of the
known decay $\psi^\prime \to J/\psi \, \pi \, \pi$,
and both amplitudes are parameterized by the linear in $q^2$ expression
(\ref{aparam}). One can of course notice that due to larger mass
difference between $\ecp$ and $\eta_c$ than between $\psi^\prime$ and
$J/\psi$ the physical region in the transition from $\ecp$ extends to
somewhat larger values of $q^2$. If the linear in $q^2$ growth of the
amplitude in eq.(\ref{aparam}) extends over those extra $75-95 \, MeV$
of the dipion invariant mass, the phase space integral for the total
decay rate is dramatically enhanced. Numerically, under this assumption
and using $C=4.6 \, m_\pi^2$ for both decays one finds after the phase
space integration that
\beq
{\Gamma(\ecp \to \eta_c \, \pi \, \pi) \over \Gamma(\psi^\prime \to
J/\psi \, \pi \, \pi)} = 3.5 \pm 0.5~,
\label{ratio}
\eeq
where the uncertainty corresponds to the experimental error $\pm 10 \,
MeV$ in the mass of $\ecp$.

The assumption that the linear in $q^2$ behavior of the amplitude
persists sufficiently beyond the physical region of the decay
$\psi^\prime \to J/\psi \, \pi \, \pi$ does not look unreasonable, given
the very high degree of linearity observed in the $\psi^\prime$ and
$\Upsilon^\prime$ decays (see a discussion in the review \cite{vyz}). In
any case, this issue would be an interesting point to study once the
discussed decay of $\ecp$ is observed experimentally.

An estimate of the relative significance of the decay $\ecp \to \eta_c
\, \pi \, \pi$ i.e. of its branching ratio is somewhat less certain
because of difficulty of a reliable estimate of the total width of
$\ecp$. Given that the width of the radiative decay $\ecp \to h_c \,
\gamma$ is rather small, and is estimated to be approximately $40-50 \,
KeV$\cite{elq}, the total width is dominated by the gluonic annihilation
rate, which can be estimated as $\Gamma(\ecp)/\Gamma(\eta_c) \approx
\Gamma(\psi^\prime \to e^+e^-)/\Gamma(J/\psi \to e^+e^-) \approx 0.4$.
However, as already mentioned, the total width of $\eta_c$ is known
quite poorly. Under these circumstances, using $4-5 \, MeV$ as a
representative value for $\Gamma(\ecp)$, one can expect that the
discussed transition with two charged pions should have the branching
ratio close to the range of 5-10\% and thus should be the exclusive
decay channel with the largest branching ratio.

This work is supported in part by the DOE grant DE-FG02-94ER40823.

\end{document}